EndHiC: assemble large contigs into chromosomal-level scaffolds using the Hi-C links from contig ends

Sen Wang[†], Hengchao Wang[†], Fan Jiang[†], Anqi Wang, Hangwei Liu, Hanbo Zhao, Boyuan Yang, Dong Xu, Yan Zhang[*], Wei Fan[*]

Guangdong Laboratory for Lingnan Modern Agriculture (Shenzhen Branch), Genome Analysis Laboratory of the Ministry of Agriculture and Rural Affairs, Agricultural Genomics Institute at Shenzhen, Chinese Academy of Agricultural Sciences, Shenzhen, Guangdong, 518120, China

[†]The authors wish it to be known that, in their opinion, the first three authors should be regarded as joint First Authors.

[*]To whom correspondence should be addressed.

**ABSTRACT**

**Motivation:** The application of PacBio HiFi and ultra-long ONT reads have achieved huge progress in the contig-level assembly, but it is still challenging to assemble large contigs into chromosomes with available Hi-C scaffolding software, which all compute the contact value between contigs using the Hi-C links from the whole contig regions. As the Hi-C links of two adjacent contigs concentrate only at the neighbor ends of the contigs, larger contig size will reduce the power to differentiate adjacent (signal) and non-adjacent (noise) contig linkages, leading to a higher rate of mis-assembly.

**Results:** We present a software package EndHiC, which is suitable to assemble large contigs (> 1-Mb) into chromosomal-level scaffolds, using Hi-C links from only the contig end regions instead of the whole contig regions. Benefiting from the increased signal to noise ratio, EndHiC achieves much higher scaffolding accuracy compared to existing software LACHESIS, ALLHiC, and 3D-DNA. Moreover, EndHiC has few parameters, runs 10-1000 times faster than existing software, needs trivial memory, provides robustness evaluation, and allows graphic viewing of the scaffold results. The high scaffolding accuracy and user-friendly interface of EndHiC, liberate the users from labor-intensive manual checks and revision works.

**Availability and implementation:** EndHiC is written in Perl, and is freely available at https://github.com/fanagislab/EndHiC.

**Contact:** fanwei@caas.cn and milrazhang@163.com

**Supplementary information:** Supplementary data are available at *Bioinformatics* online.

**1 Introduction**

The long accurate reads generated by PacBio HiFi technology, with an average read length over 10-Kb and base accuracy over 99%, has been widely applied in *de novo* assembly of plant and animal genomes (Marx, 2021). Specific software for assembling HiFi reads has also been developed, such as Hifiasm (Cheng, et al., 2021) and HiCanu (Nurk, et al., 2020), which in general can produce large contigs with N50 size over tens of mega-bases for most genomes. Besides, the ultra-long noisy reads with an average length over 50-Kb from Oxford Nanopore Technology (ONT) and related software development such as wtdbg2 (Ruan and Li, 2020), also largely improve the contig continuity in genome assemblies. The huge progress in the contig-level assembly makes the goal of telomere-to-telomere assembly more visible than ever before, however, chromosomal-scale linkage data is still needed to assemble these large contigs into complete chromosomes.

Hi-C sequencing provides a fast and cost-effective way for scaffolding the contigs into chromosomal-level sequences (Lieberman-Aiden, et al., 2009), based on the facts that the density of Hi-C linked pairs between adjacent contigs is higher than that of non-adjacent contigs. Several Hi-C scaffolding software have been developed: LACHESIS and ALLHiC use hierarchical agglomerative clustering algorithm to group contigs, and perform ordering and orientation in the following steps (Burton, et al., 2013; Zhang, et al., 2019); 3D-DNA cuts contigs in the middle site to form sister-contigs, assigns links to non-sister contigs with reciprocal best requirement, and finishes clustering, ordering and orientation tasks simultaneously (Dudchenko, et al., 2017); HiC-Hiker aims to improve the contig order and orientation generated by other Hi-C scaffolding tools, adopting a probability model and dynamic programming algorithm (Nakabayashi and Morishita, 2020); HiRise is specially designed for *in vitro* Hi-C data (Putnam, et al., 2016), while SALSA2 constructs scaffolds using both edges from the contig GFA (Graphical Fragment Assembly) file and linkages inferred from the Hi-C data (Ghurye, et al., 2019). However, with all available software, it is still challenging to correctly assemble large contigs into chromosomes, and manual checking and revising are often necessary.

Logically, large contigs will simplify scaffolding with Hi-C data, but it is not as expected in practice. To the best of our knowledge, all available software computes the pairwise contig linkages using Hi-C links from the whole contig regions (Burton, et al., 2013; Dudchenko, et al., 2017; Zhang, et al., 2019). In this article, our analysis showed that the Hi-C links of two adjacent contigs concentrate only at the neighbor ends of the contigs (usually < 3-Mb), and large amount Hi-C links also exist between some non-adjacent contigs. As the contig size gets larger, the number of Hi-C links between adjacent contigs and that between non-adjacent contigs will become closer. Therefore, the large contig size will reduce the power to differentiate adjacent contig linkages from non-adjacent contig linkages calculated using Hi-C links from the whole contig regions, leading to a high error rate in scaffolding these large contigs. To resolve this problem, we developed a new Hi-C scaffolding tool EndHiC, constructing scaffolds with only the Hi-C links from contig ends, which is more suitable for the assembling of large contigs (> 1-Mb) into chromosomes.

**2 Description**

The overall workflow of EndHiC is shown in Figure 1a, and EndHiC needs two input data: the high continuous and accurate contigs generated by Hifiasm (Cheng, et al., 2021), HiCanu (Nurk, et al., 2020), etc, and the analyzing results of reads mapping from HiC-Pro pipeline. To fully utilize the professional efforts of HiC-Pro (Servant, et al., 2015), the ".bed" file recording relationships between contigs and bins, the raw and normalized ".matrix" files storing contact values of all bin pairs, with bin size of 100-Kb (recommend), are used by EndHiC. The contact value for each pair of contig ends is calculated by summing the contact values of bins falling into the corresponding contig end regions. Each contig has two ends, the left-side end is defined as head end, while the right-side end is defined as tail end. The size of the contig ends is determined as the number of 100-Kb bins (Figure 1b).

The size of the contig ends is vital for the effectiveness of EndHiC. Then, what is the optimal size value? To answer this question, we calculated the contact values for contig ends with different contig end sizes ranging from 500-Kb to 40-Mb, using the great burdock (*Arctium lappa*) genome data, which is finely assembled by Hifiasm with all the large contigs correctly scaffolded into 18 chromosomes. We took the contact values from adjacent contig ends as signal contact values, while that from non-adjacent contigs ends as noise contact values. To evaluate the power of using various contig end sizes to differentiate between signal and noise contact values, we adopted a signal to noise ratio (SNR) approach. For each contig end size, the SNR is defined as the median of signal contact values divided by the median of noise contact values. From the distribution of SNR along contig end sizes, we observed that the SNR is higher than 100 : 1 for contig end size less than 2.5-Mb, but becomes lower than 10 : 1 for contig end size larger than 20-Mb (Figure 1c & Figure S1). That is to say, using the contig end sizes larger than 2.5-Mb will have lower ability to differentiate between adjacent and non-adjacent contig linkages. On the other hand, we also noticed that the number of Hi-C links fluctuate severely within smaller contig regions. Thus, the use of contig end sizes smaller than 500-Kb tends to generate instable scaffold results. To sum up, contig end sizes from 500-Kb to 2.5-Mb are recommended for EndHiC.

When the contact values among all the contig ends with a specified size are calculated, the key job is to differentiate the signal from noise contact values, which are generated from adjacent and non-adjacent contig ends, respectively. Is there a cutoff value able to clearly separate the signal and noise contact values? To resolve this question, we sorted all the contact values of contig ends (1.5-Mb) in great burdock, and found a sharp turning point around 1,500 in the distribution (Figure S2a). The contact values larger than the turning point composes all the signal contact values and a small number of noise signal values, whereas the contact values smaller than the turning point composes only the noise contact values. Based on this finding, we developed a linear transforming method to automatically detect the turning point (Figure S2b-c), which is used as a basic contact cutoff in EndHiC. In addition, EndHiC also adopts a reciprocal best approach, requiring the contact value between two adjacent contig ends to be the max contact value for both of the contig ends, which further filters some noise contact values larger than the turning point.

A reliable link will be assigned to two contig ends, if they have a contact value over the basic cutoff and also satisfying reciprocal best. Using the contigs as nodes, and the reliable links among contigs as edges, a graph structure is constructed. Then, EndHiC converts the graph structure into a GFA format file, following the rules showed in Figure 1d. If the link is assigned to the tail end of contig_1 (on the left side), a strand mark "+" will be given to contig_1, whereas if the link is assigned to the head end of contig_1, a strand mark "-" will be given to contig_1; the opposite rules are applied to contig_2 (on the right side). Then, the GFA file can be viewed in Bandage, a tool for visualizing assembly graphs with connections (Wick, et al., 2015). As an example, the Bandage view of great burdock scaffolds generated by EndHiC is shown in Figure 1e. With the graphic tool, we can quickly have an overall view about the scaffolding quality, and find the assembly errors easily. Given the reciprocal best requirement, one contig end will have only one link to the other contig end, which makes sure that most paths in the GFA file are linear. However, circular path might still occur when the two ends of a chromosome are connected. EndHiC firstly breaks the circular paths at the position with lowest contact value, and then generates clusters with order and orientation information from all the linear paths.

To increase scaffolding accuracy, the default EndHiC pipeline runs in parallel with various contig end sizes from 500-Kb to 2500-Kb in step of 500-Kb, and various contact cutoffs from 1 to 5 times of the basic cutoff (turning point value) in step of 0.5, using both raw and normalized matrix data. Each combination of the parameters generates an independent cluster file with contig order and orientation. Then, EndHiC summarizes all these cluster results (Table S1), merges the cluster units and counts frequency. Cluster units with a relatively high frequency are thought as stable clusters, and a set of stable clusters without redundant contigs is presented as the output, which is considered as the most accurate result of EndHiC. The frequency of the cluster unit is also used as a robustness evaluation, reflecting the accuracy of each individual cluster (Table S2). Finally, EndHiC converts the stable cluster file into AGP and Fasta format files, which are compatible for most downstream analysis.

The performance and accuracy of EndHiC was compared with LACHESIS, ALLHiC, and 3D-DNA, using both the simulated contig data (human, rice, Arabidopsis) and the real contig data (human, great burdock, water spinach) (Table S3). All the Hi-C data used here are real sequencing data, with coverage depth ranging from 40 to 130 X (Table S4). One default round of EndHiC is applied to all the datasets except for the human real data, which needs two rounds of EndHiC. In the human real data, some contig ends have relatively lower link density due to long heterochromatin repeat sequences, so the first round of EndHiC failed to cluster these contigs into chromosome-level scaffolds. Larger contig ends will span the repetitive regions in the contig end and overcome this problem, so the second round using contig end sizes 3.0-Mb to 5.0-Mb in step of 0.5-Mb have clustered these contigs. Generally, for most genomes, only a few minutes and less than 1-G memory is needed by EndHiC, which runs ~10 times faster than LACHESIS and ALLHiC, ~1000 times faster than 3D-DNA, and consumes less than 1/100 memory than the other three software (Table S5).

We used the corresponding reference genomes to validate the assembly accuracy of species human, rice, and Arabidposis, and used the Bandage graphic view (Figure S3), Hi-C heatmap (Figure S4), as well as the synteny relationship with closely related species (Figure S5), to validate the assembly accuracy of species great burdock and water spinach, whose reference genome are not available. EndHiC correctly clustered all the simulated and real large contigs into chromosomes, in contrast, 3D-DNA, ALLHiC, LACHESIS only correctly clustered 50%, 15%, 0% of the chromosomes, respectively, in average of the simulated and real data (Table S6). In respect to order and orientation, EndHiC only makes a mistake for two neighbor contigs in chromosome_1 of the human real data, caused by the extremely long heterochromatin repeat sequence (~20 Mb) between them. Other software can't even cluster these two contigs together, not to mention the ordering and orientation. In summary, these results showed that EndHiC has much higher accuracy in clustering and comparable or higher accuracy in ordering and orientation of large contigs, compared to other available software.

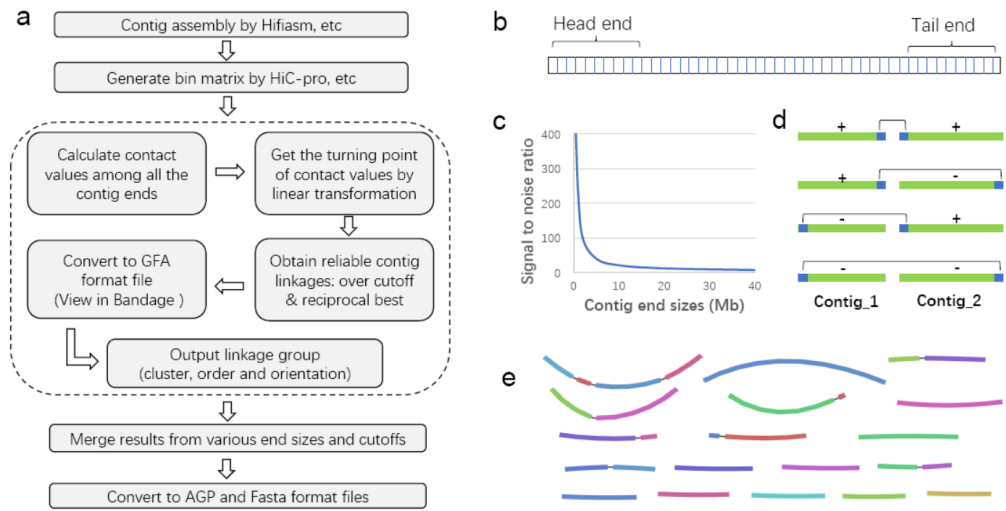

**Figure 1.** Illustration of EndHiC; a. The overall workflow of EndHiC, the five parts inside the dashed lines are the core of EndHiC; b. Schema chart for contig head and tail end with size 1-Mb, each end contains 10 of 100-kb bins; The last bin of the contig is generally shorter than 100-Kb, but for bin number over 5, i.e. contig end size over 500-Kb, suggested in practice, the size of head and tail ends using specified number of bins can be thought as equal. c. Distribution of signal to noise ratio (SNR) along contig end sizes, calculated from the raw links data of great burdock. d. Rules for determining relative contig order and orientation by the reliable links between neighbor contig ends. The color green and blue indicate contig and contig end, respectively. The connecting line indicate reliable link between adjacent contig ends. The mark "+" and "-" means forward and reverse strands assigned in the GFA file. e. Example graphic view (Bandage) of EndHiC result for great burdock, 28 large contigs were assembled into 18 large scaffolds (anchor rate >99%), each corresponding to a natural chromosome, small clusters are not shown. Using raw and normalized matrix data generates the same result.

## 3 Discussion
EndHiC implements a simple but effective Hi-C scaffolding algorithm, leveraging on HiC-Pro professional works for valid pair detection, calculation and normalization of bin matrix. Although not designed to be a universal Hi-C scaffolding tool, EndHiC is applicable to large and accurate contigs. The algorithm of EndHiC is more similar to 3D-DNA than other available tools, as the contig head and tail ends in EndHiC has similar function to the sister-contigs in 3D-DNA. Besides, neither EndHiC nor 3D-DNA needs the chromosome number as input, and they both perform clustering, ordering, and orientation simultaneously. Unlike 3D-DNA, EndHiC does not integrate a module to pre-process the mis-assembled contigs, as mis-assembly in large contigs becomes fewer. If necessary, the users can manually break the erroneous contigs according to the Hi-C heatmap of contigs, or invoke the error correction module of 3D-DNA or other software. By using the Hi-C links from contig ends, EndHiC has greater SNR compared to other available software, contributing to higher accuracy of the scaffold assembly. The scaffolding results from using various contig end sizes and various contact cutoffs with both raw and normalized links data, have also been compared and merged to get a more accurate assembly as well as a robustness evaluation. For genomes with a high proportion of heterochromatin, long repetitive sequences often exist in the contig ends, where one default round of EndHiC may can't finish clustering all the large contigs into chromosome-level scaffolds. To resolve this problem, we suggest running EndHiC multiple rounds, each round with increasing contig end sizes, until the resulting cluster number is equal or close to the known or estimated chromosome number.

For most plant and animal genomes, the contigs derived from current sequencing and assembling technologies might consist of only a very small ratio (< 1%) of small contigs (< 1-Mb), most of which are rich in repeat sequences and have low Hi-C link density. These contigs are ignored by EndHiC firstly, because they don't meet the minimum contig size requirement that should be at least larger than two times of the used contig end size (500-Kb). By removing the interference of small repetitive contigs prone to cause mis-assemblies, it also assures the assembly accuracy of those large contigs. Then, the EndHiC package provides a program to cluster these small contigs into the constructed chromosome-level scaffolds based on the overall maximum Hi-C links. Compared to large contigs, the order and orientation for small contigs is more difficult to determine, and our contig end rule no longer works. We suggest to borrow from the global optimization algorithm used in HiC-Hiker (Nakabayashi and Morishita, 2020) or the genetic algorithm used in AllHiC (Zhang, et al., 2019), to obtain acceptable order and orientation for these small contigs.

**4 Conclusion**
We designed and implemented a fast and memory-efficient Hi-C scaffolding tool EndHiC, using Hi-C links from the contig end regions instead of the whole contig regions. EndHiC is suitable to assemble large contigs into chromosomal-level scaffolds, and has much higher accuracy compared with existing Hi-C scaffolding tools. Moreover, EndHiC runs extremely fast with trivial memory consumption, and is easy to use, with no parameter needed to be set by the users at most cases. In summary, the reliable cutoff of contact values, the reciprocal best requirement, the GFA graphic viewing in Bandage, and the robustness evaluation, ensure high scaffolding accuracy, liberating the users from labor-intensive manual check and revision works.

**Acknowledgements**
We thank Xingtan Zhang, Weihua Pan, and Jue Ruan for helpful discussions and constructive suggestions.

**Funding**
This work was supported by National Natural Science Foundation of China (Grant No. 32000408); Shenzhen

science and technology program (JCYJ20190814163805604); Shenzhen Science and Technology Program (Grant No. KQTD20180411143628272); Fund of Key Laboratory of Shenzhen (ZDSYS20141118170111640); The Agricultural Science and Technology Innovation Program.

*Conflict of Interest*: none declared.

## Supplementary Figures and Tables

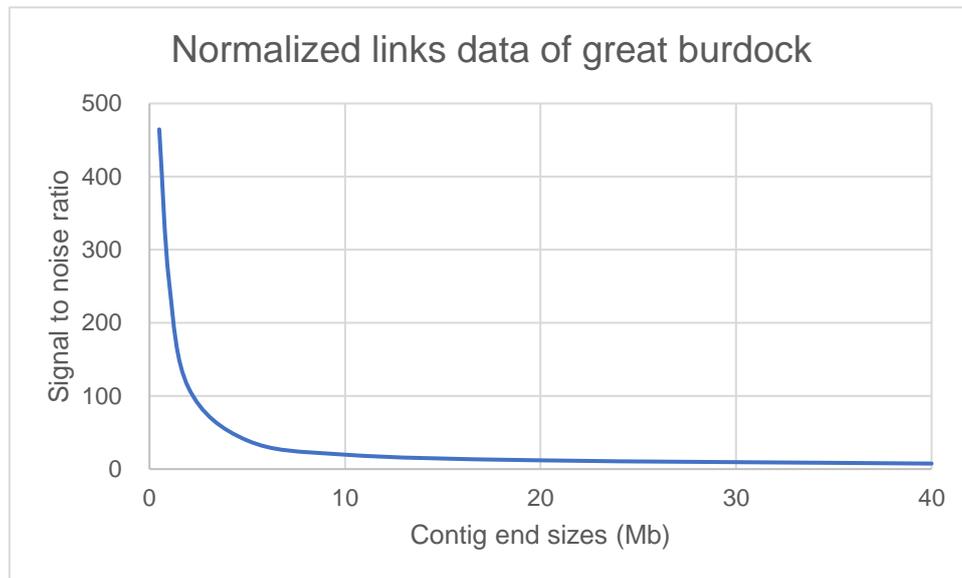

**Figure S1.** Distribution of signal to noise ratio (SNR) along contig end sizes, calculated from the normalized links data of great burdock. The contact values from adjacent contig ends and non-adjacent contigs ends are taken as signal and noise contact values, respectively. The signal to noise ratio, is defined as the median of signal contact values divided by the median of noise contact values.

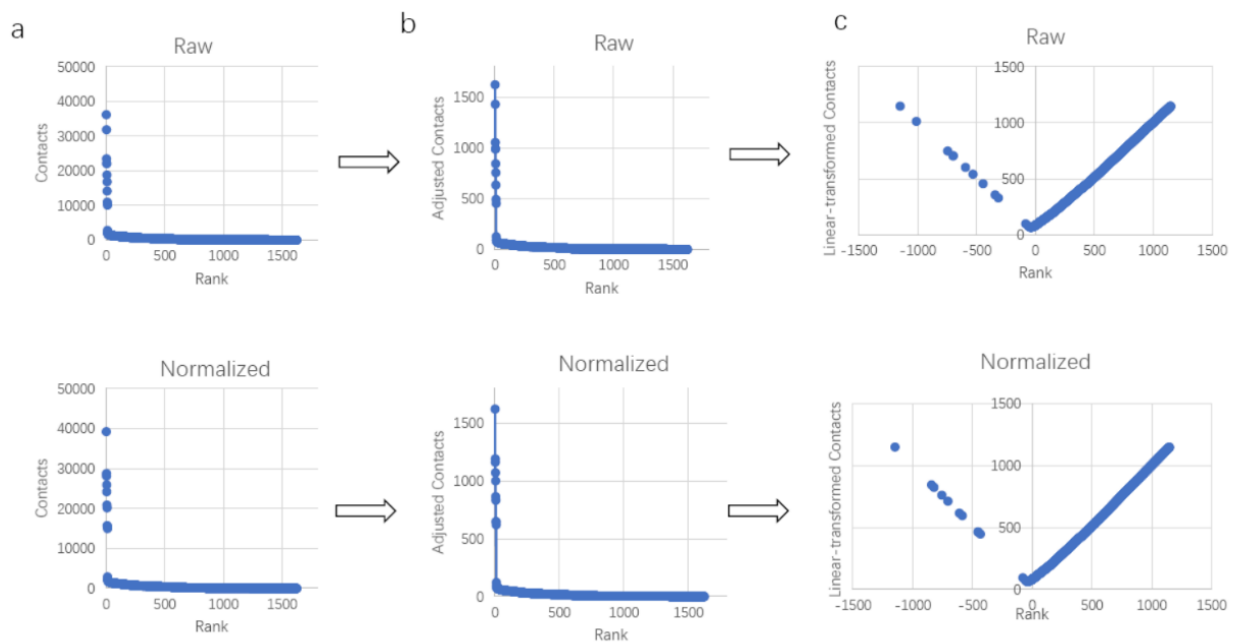

**Figure S2.** Automatic detection of the turning point in sorted contact values of contig ends. Top part uses raw links from HiC-Pro, bottom part uses normalized links from HiC-Pro. a. Distribution of sorted contact values, from max to min. b. Distribution of adjusted contact values, the scale of Y-axis is adjusted to be equal to that of X-axis. c. Distribution of linear-transformed (anticlockwise rotate 45 degree) contact values. The lowest point in the transformed data is much easier to be identified, which is equivalent to the turning point in the original data.

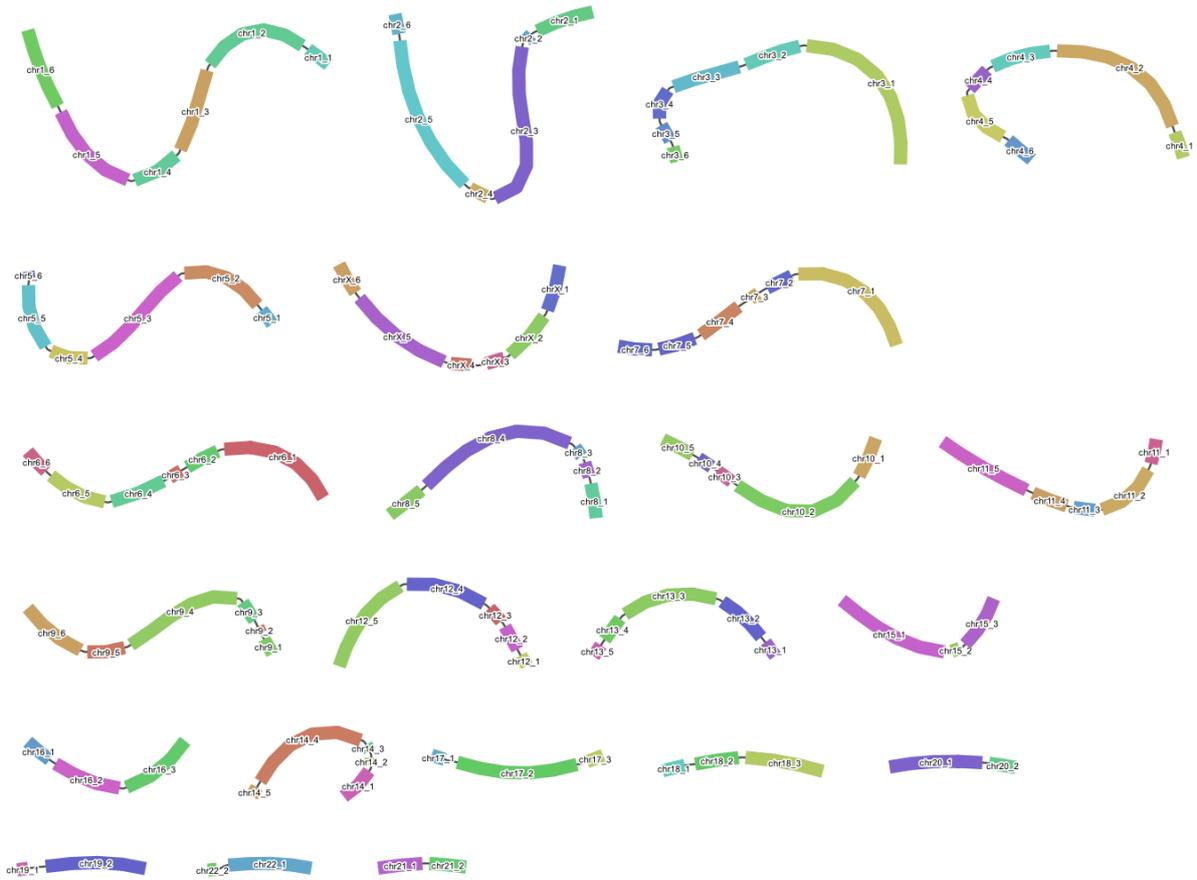
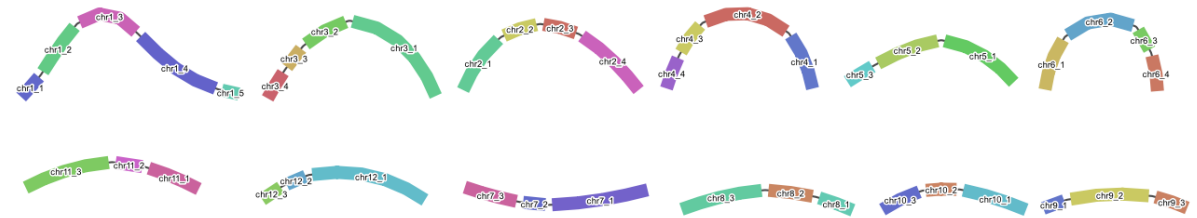
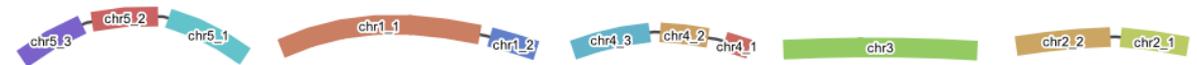

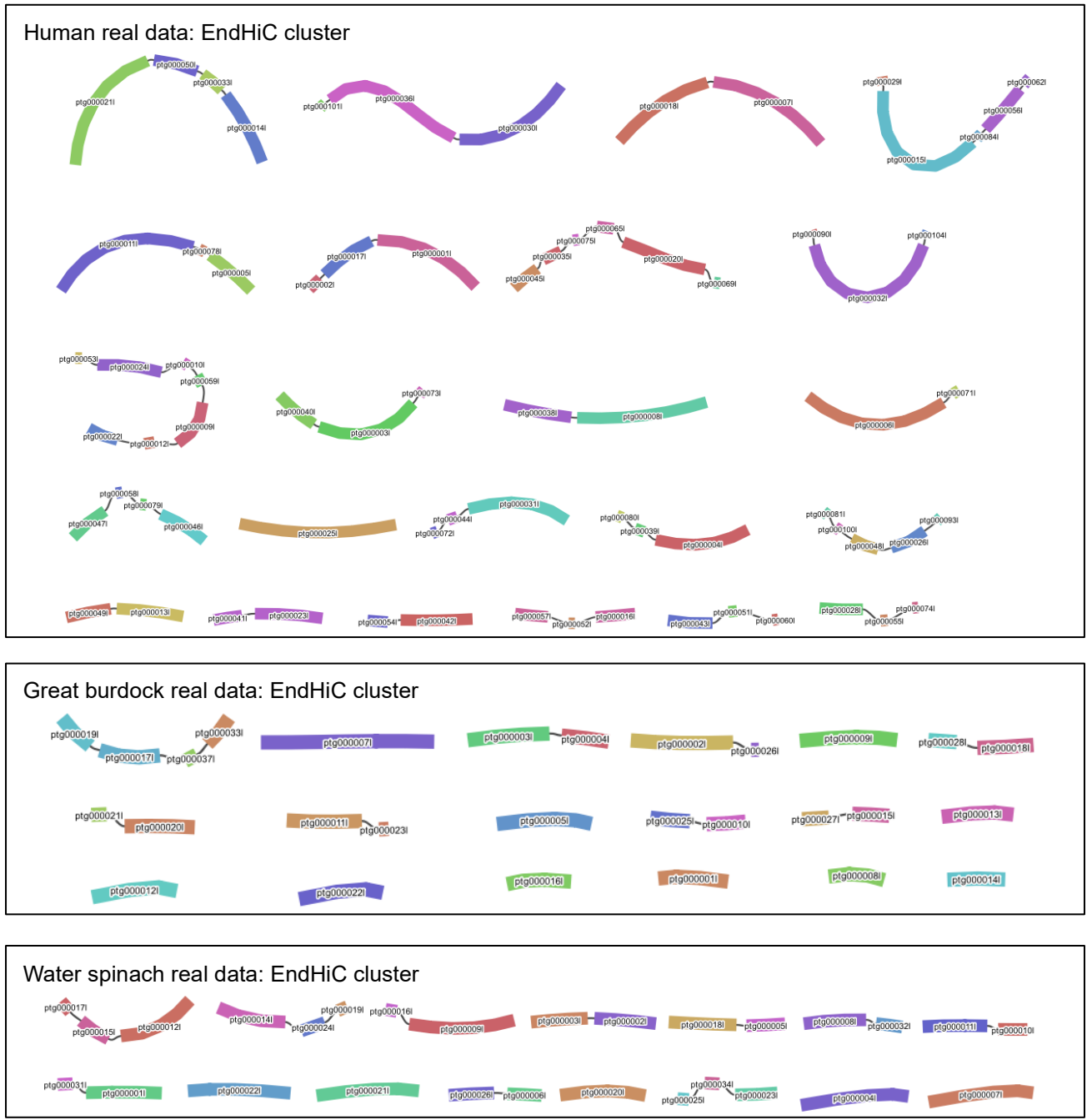

**Figure S3.** Bandage view of the final merged chromosome-level scaffolds (clusters) including contig order and orientation produced by running EndHiC on three simulated (human, rice, Arabdopsis) and three real (human, great burdock, water spinach) datasets. In the simulated data, each chromosome of reference genome was randomly split into 1 to 6 contigs, and EndHiC has anchored 100% of the contigs into chromosomes; while in the real data, the contigs were assembled by hifiasm with real PacBio HiFi reads, and EndHiC has anchored 99.3%, 99.8%, 99.7% of contigs into chromosomes for human, great burdock, and water spinach, respectively. Note that all the used Hi-C sequencing reads were real data downloaded from public database.

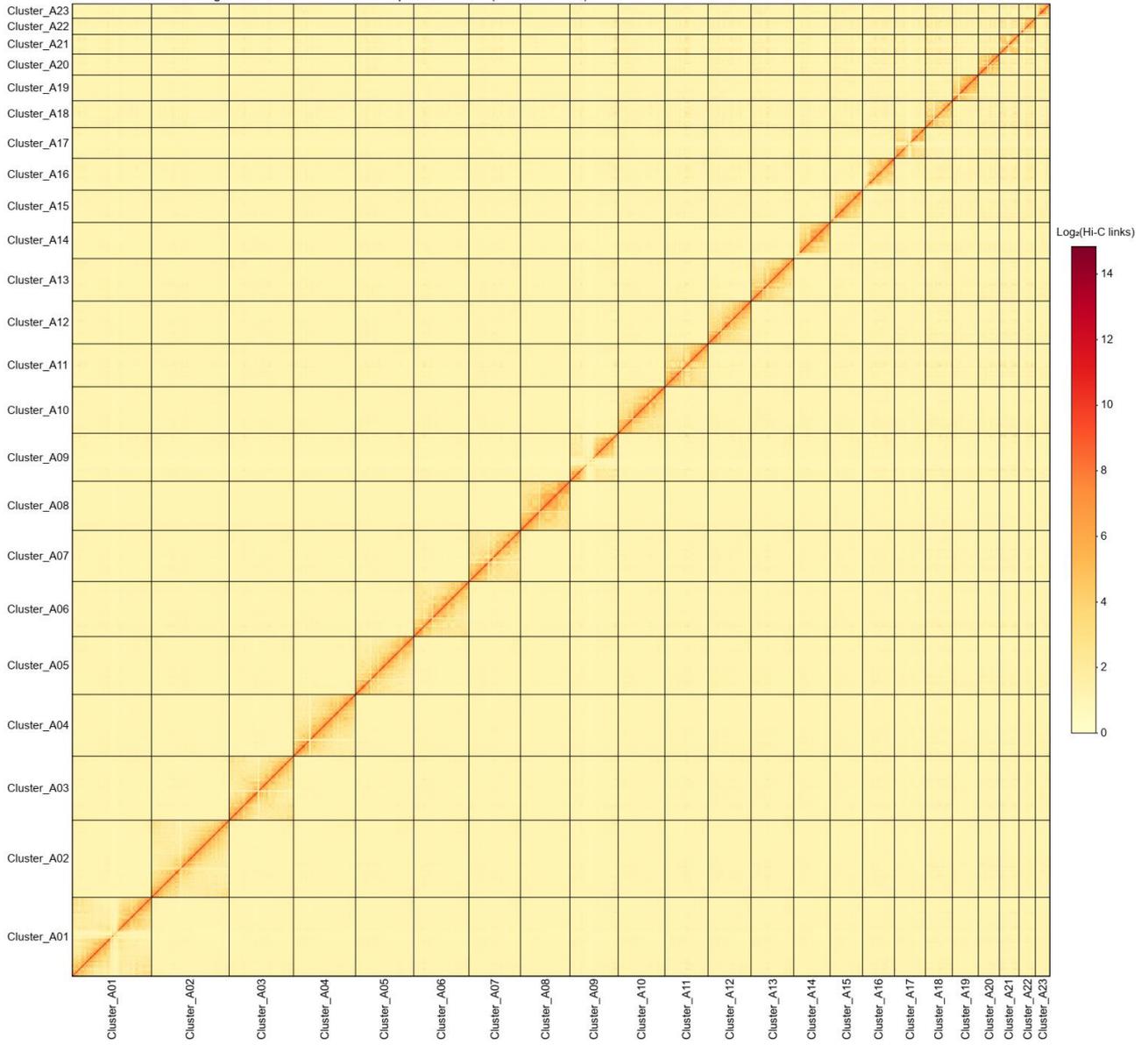

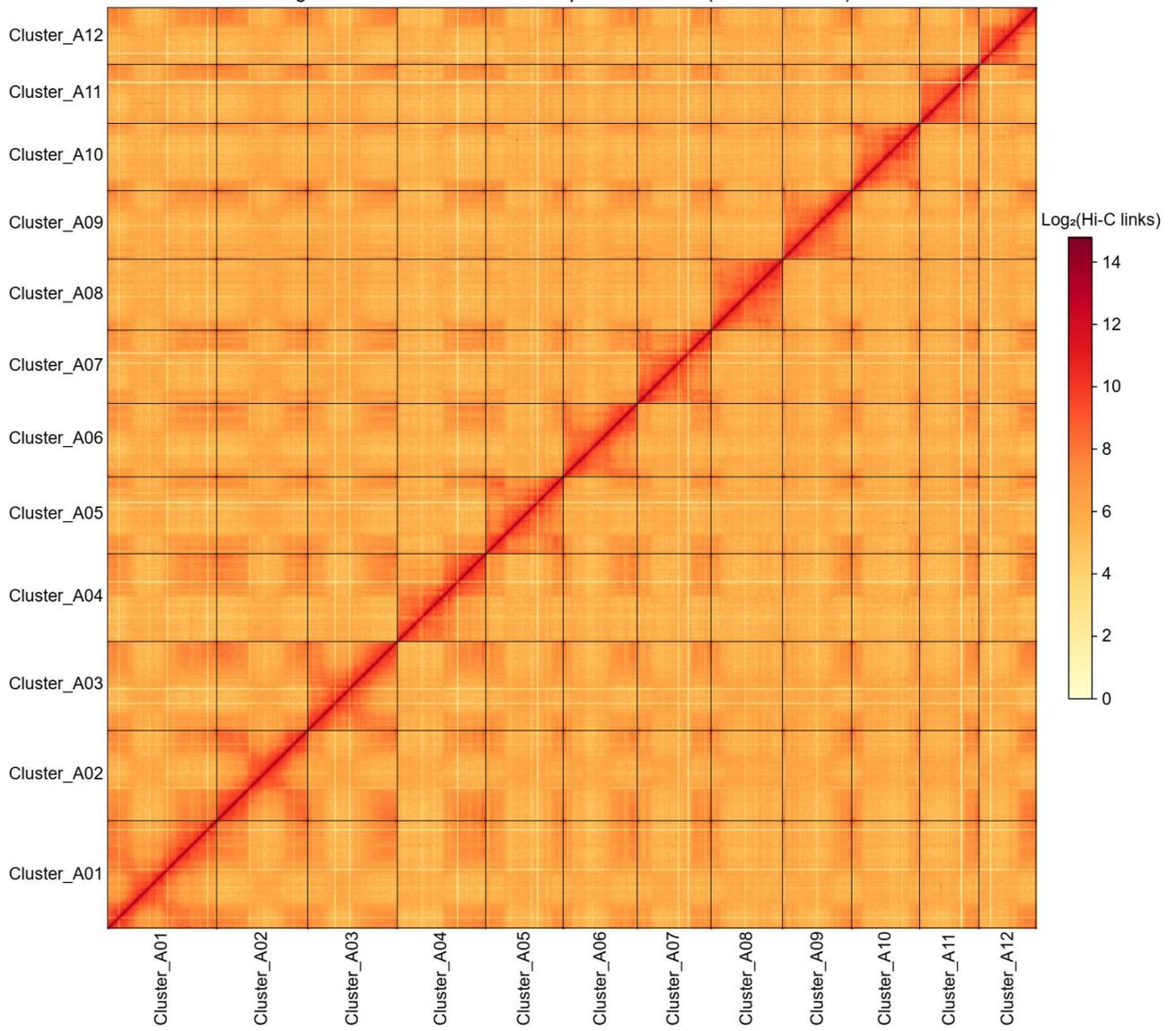

Rice simulated data: genome-wide Hi-C link heatmap of raw matrix (bin size 500Kb)

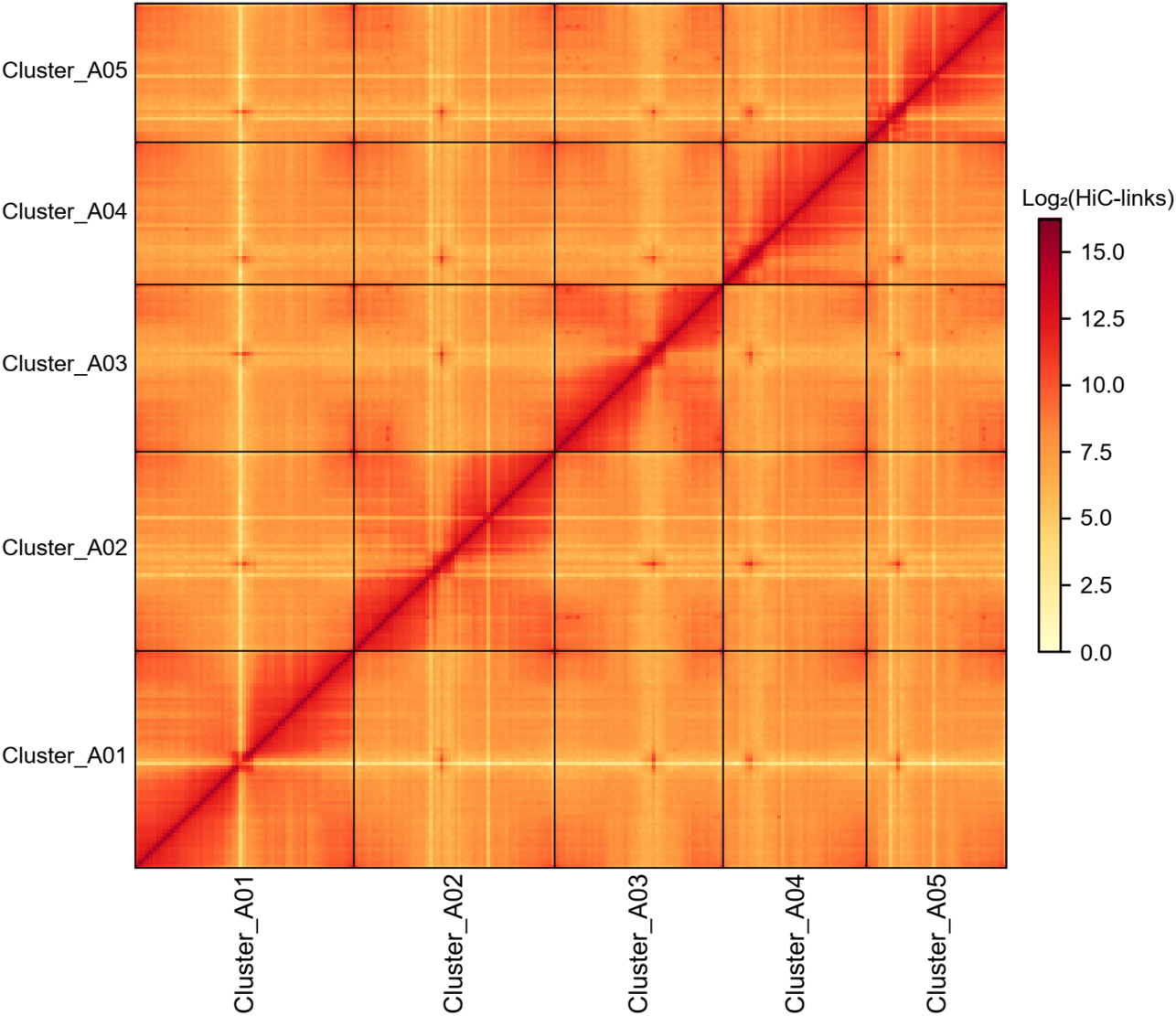

Human real data: genome-wide Hi-C link heatmap of raw matrix (bin size 500Kb)

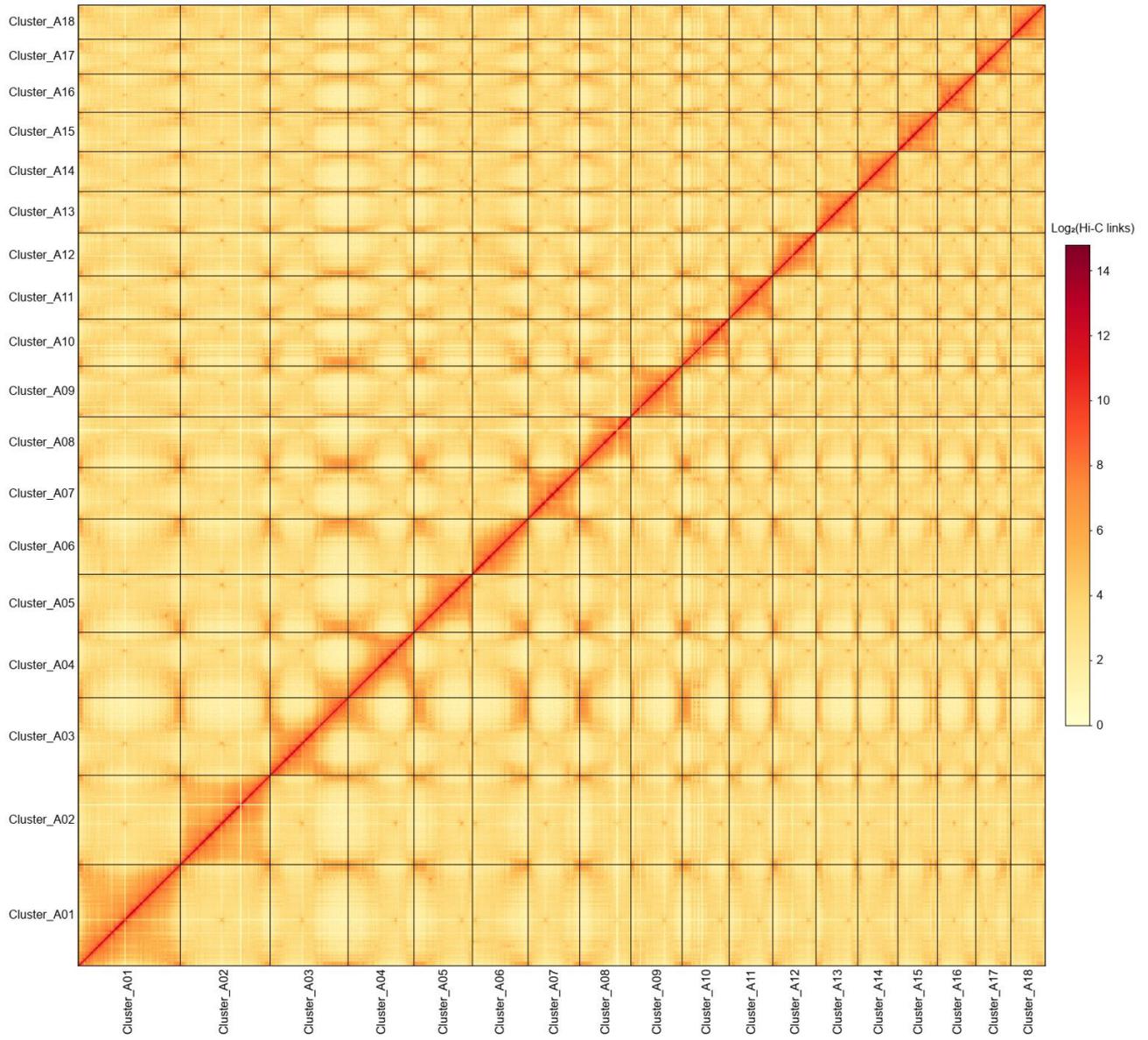

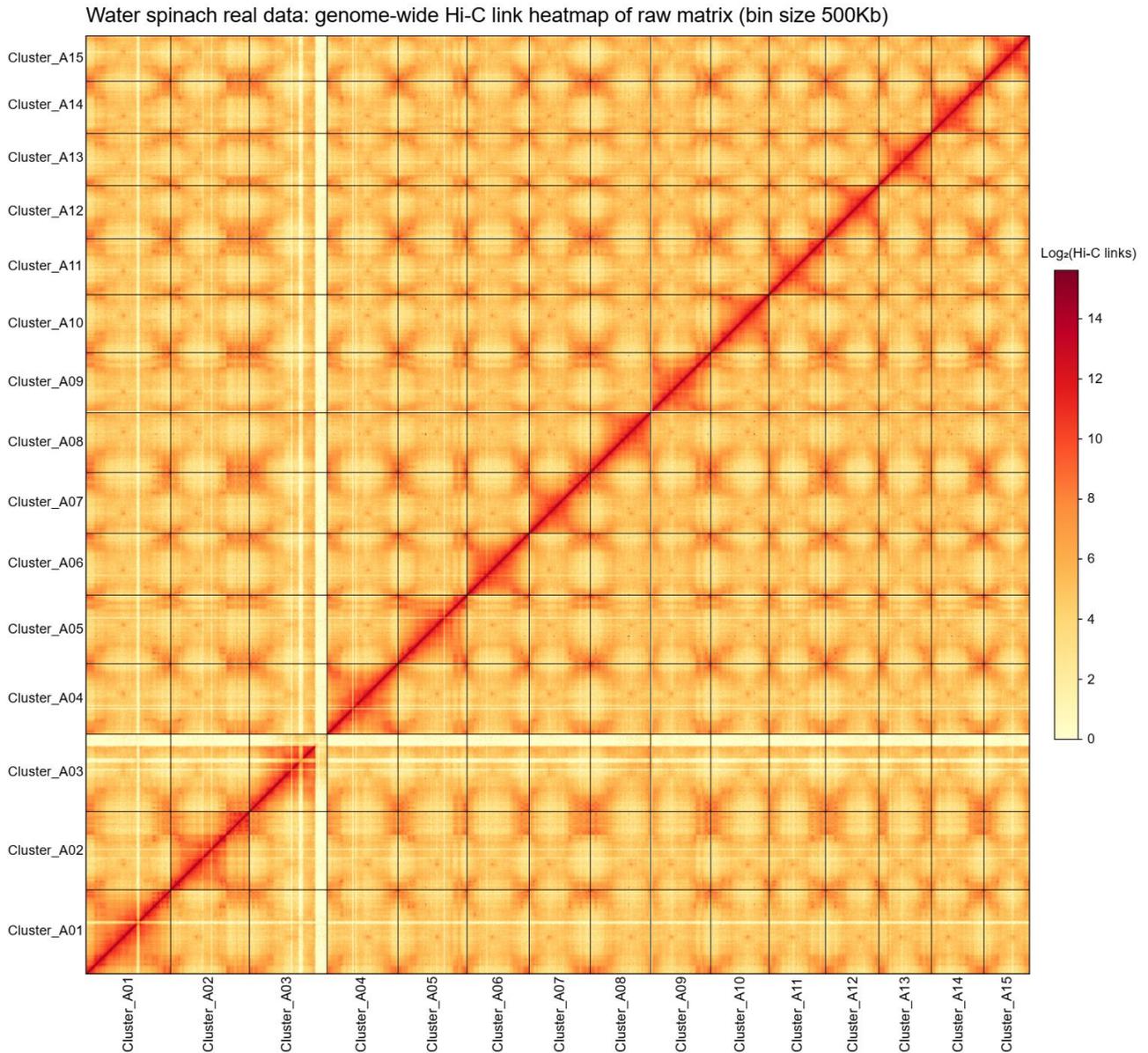

**Figure S4.** Genome-wide Hi-C link heatmap of the final merged chromosome-level scaffolds (clusters) produced by running EndHiC on three simulated (human, rice, Arabdopsis) and real (human, great burdock, water spinach) datasets. In the heatmap, each scaffold refers to a pseudo-chromosome, each pixel refers to a 500-Kb bin, and the color value indicates the base 2 logarithm of the number of valid read pairs (log2 [valid read pairs]). The plant genomes (rice, Arabidopsis, great burdock, water spinach) have more inter-chromosome Hi-C interactions than the human genome. In overall, the Hi-C link heatmap style of the simulated datasets is similar to that of real datasets.

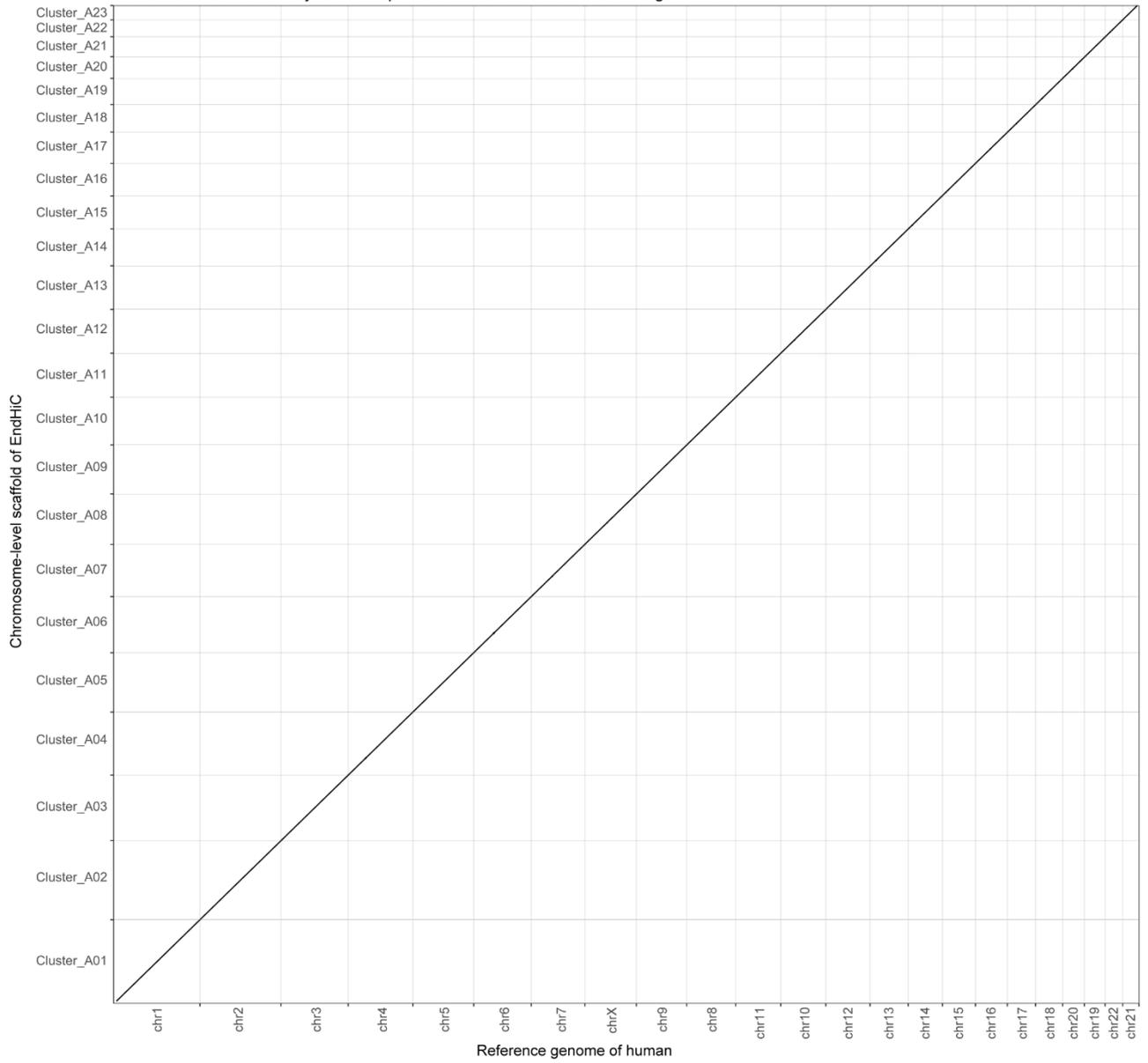

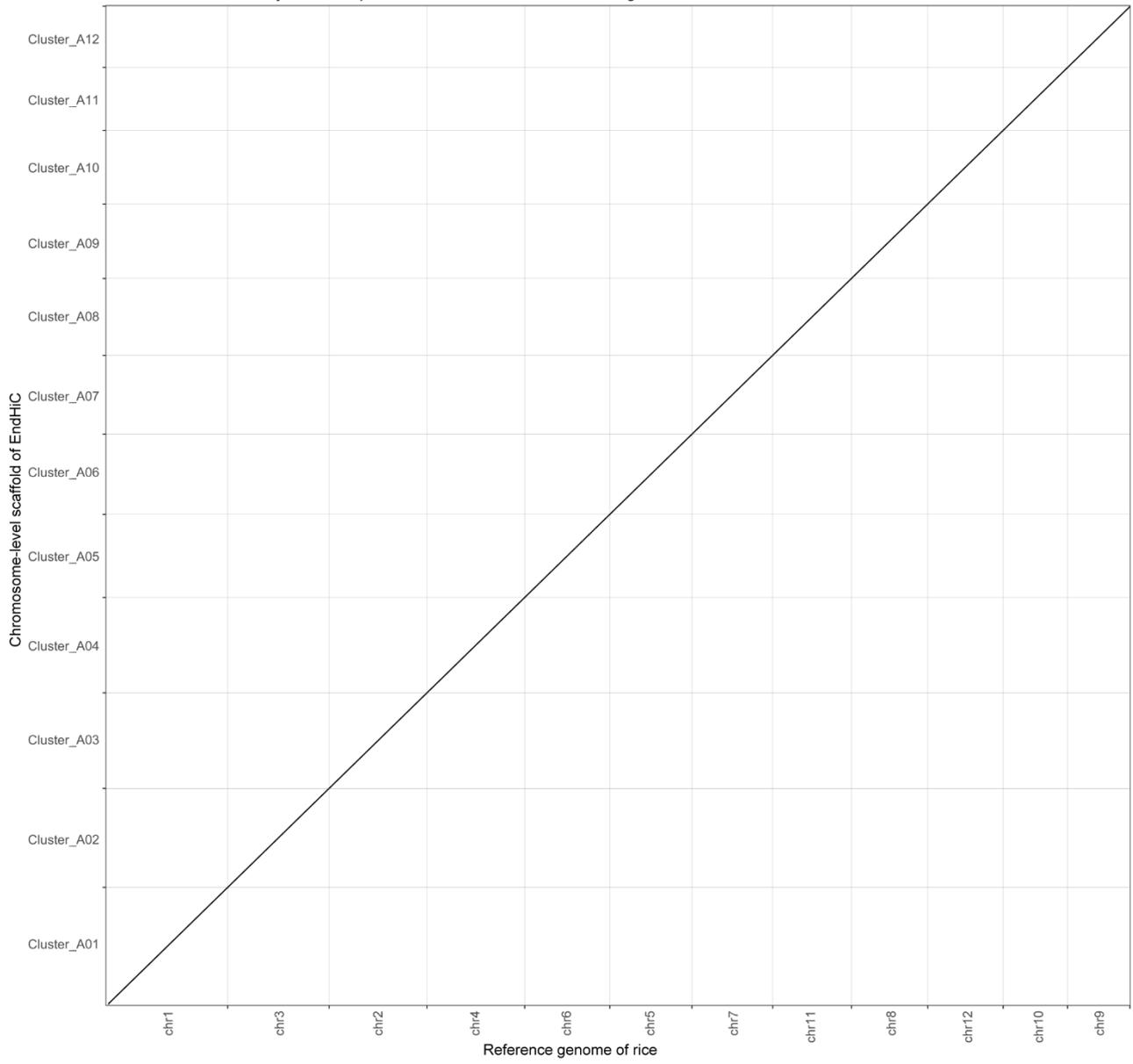

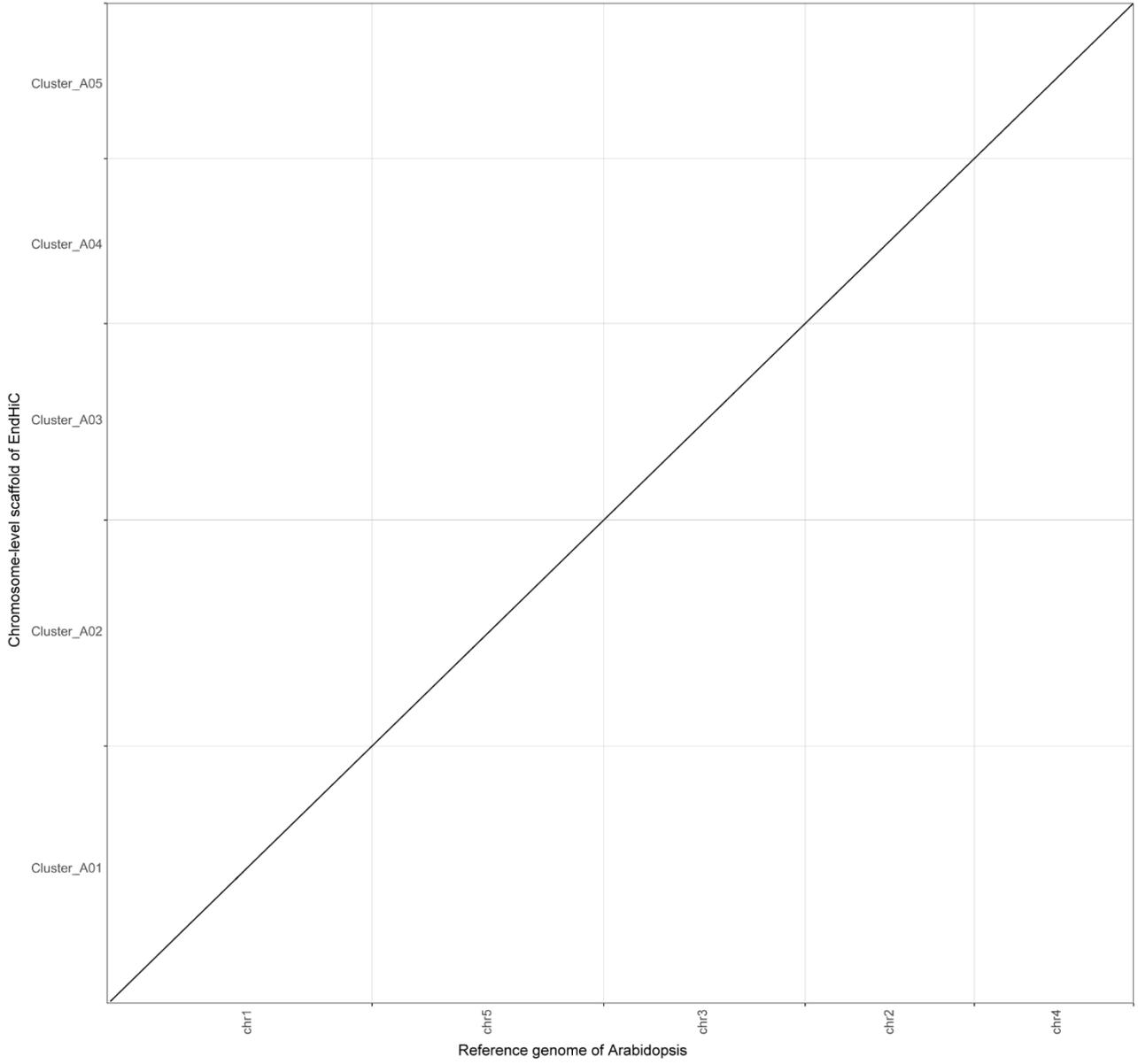

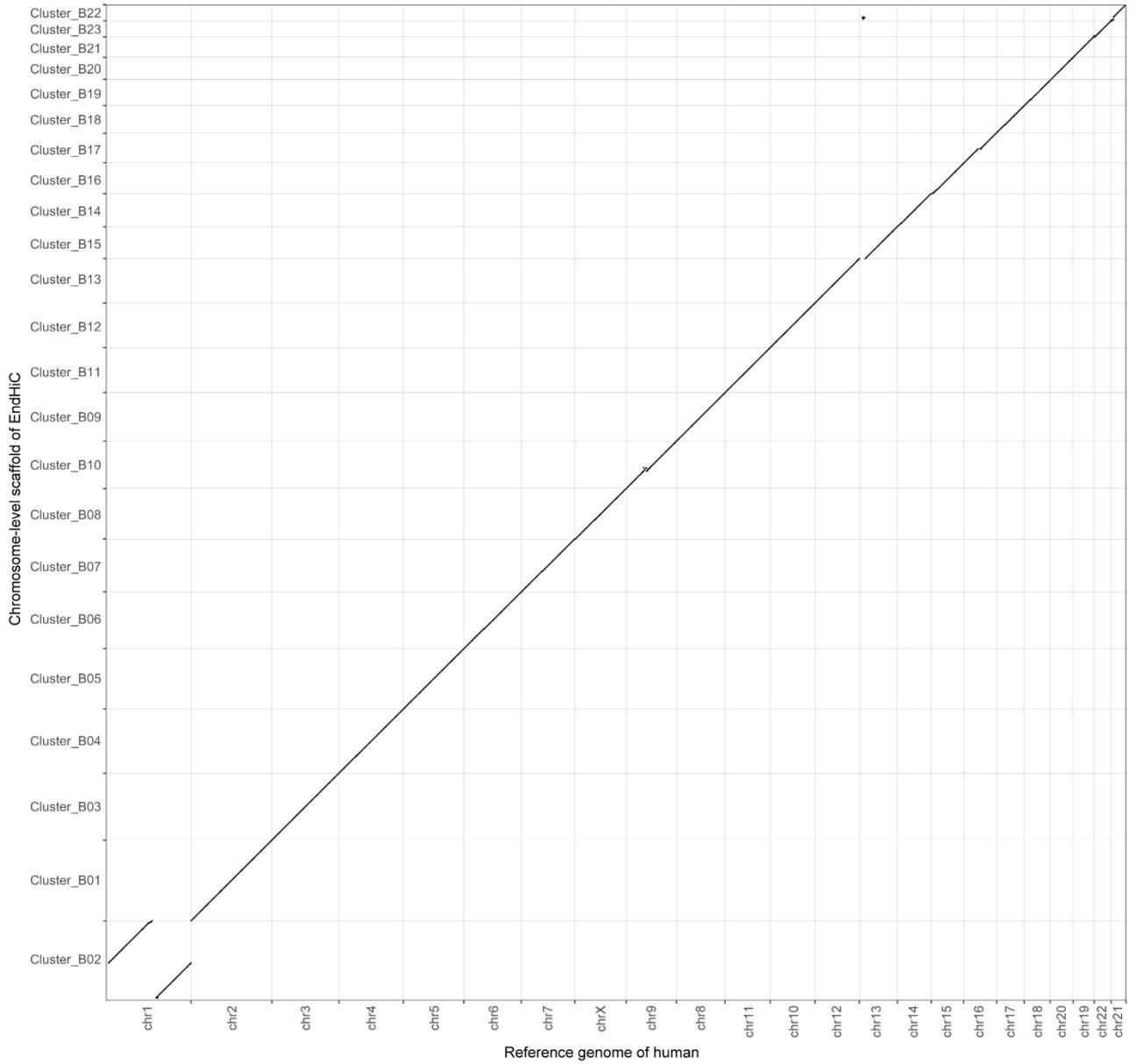

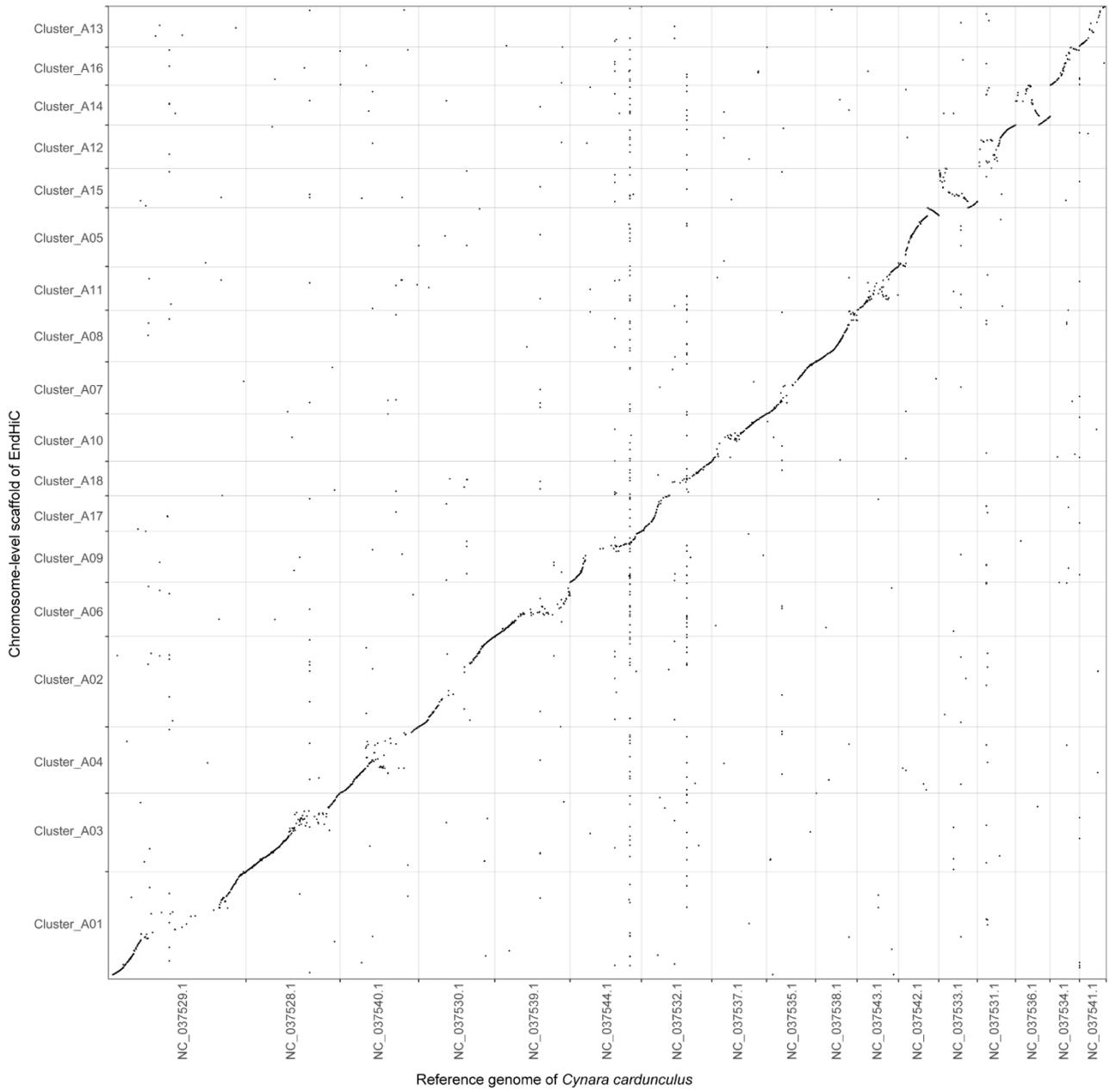

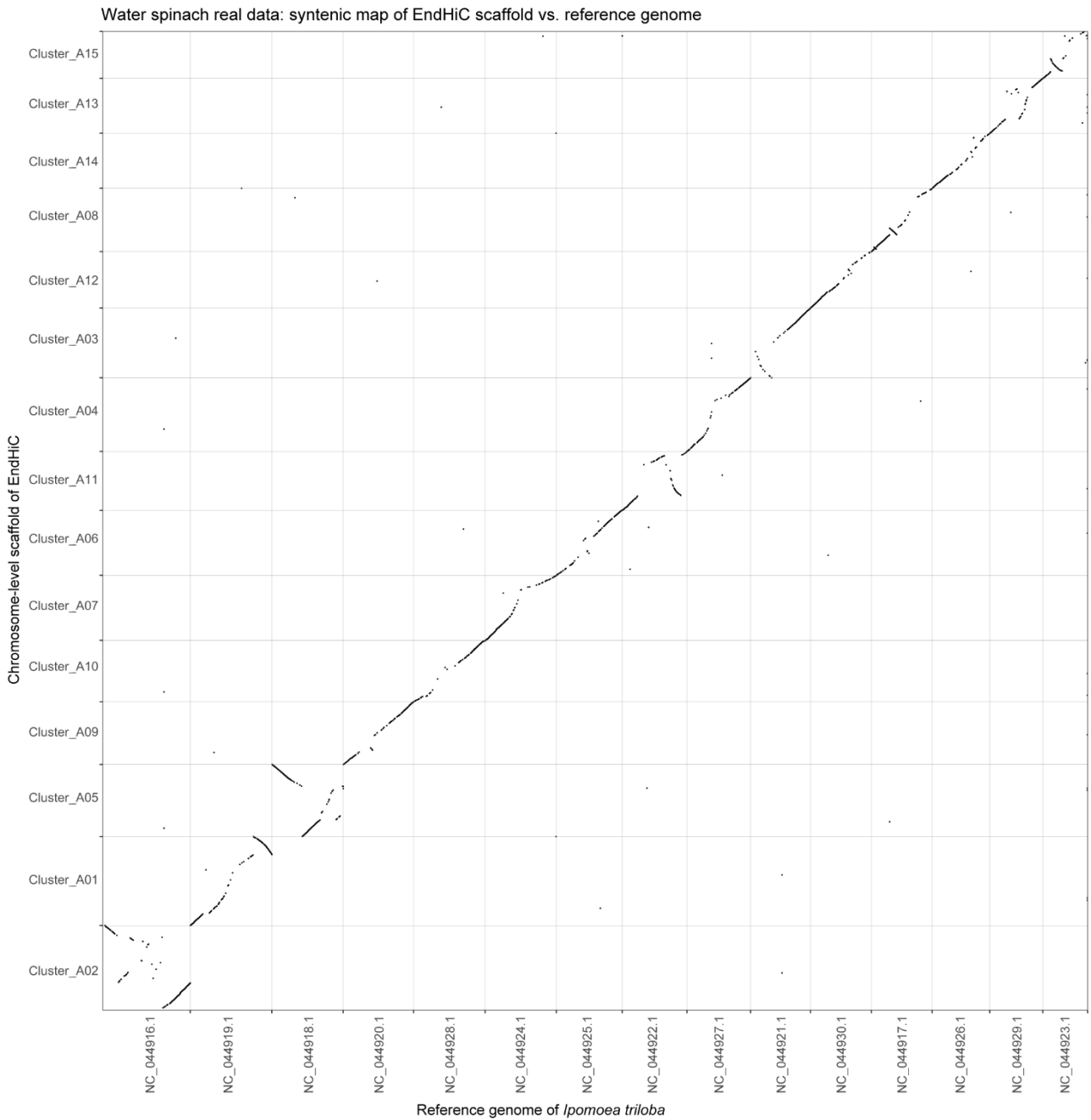

**Figure S5.** Syntenic map showing the whole-genome alignments of the final merged chromosome-level scaffolds (clusters) produced by running EndHiC to the reference genomes. The results for three simulated datasets (human, rice, Arabdopsis) and three real datasets (human, great burdock, water spinach) were shown. The corresponding reference genome of the same species (human, rice, Arabidopsis) were used for comparison, or the reference genome of closely related species (*Cynara cardunculus* and *Ipomoea triloba*) were used if the reference genome of the same species is not available. Note that great burdock has 18 chromosomes (haploid) and *C. cardunculus* has 17 chromosomes, 16 chromosomes of the former and 16 of the latter are one-to-one, except that Cluster_A17 and Cluster_A18 of burdock map to the same chromosome (NC_037532.1) of *C.cardunculus*; water spinach and *I. triloba* both have 15 chromosomes which are mapped in one-to-one manner. The whole genome alignment was performed by minimap2, and then the synteny map was drawn by dotPlotly (https://github.com/tpoorten/dotPlotly). The simulated datasets all have perfect syntenic relationships with the reference genomes. For the human real dataset, only two contigs of chromosome 1 is mis-ordered, due to the extremely long heterochromatin (~20 Mb) between them. For the real datasets of great burdock and water spinach, the syntenic relationships with their closely related species is not perfect, but largely consistent, which is reasonable considering their divergence time.

**Table S1.** Number of clusters under each parameter combination.

| Contig end size | Type of Hi-C link matrix | Times of the turning point as contact cutoff for signal and noise linkages | | | | | | | | |
|---|---|---|---|---|---|---|---|---|---|---|
| | | 1.0 | 1.5 | 2.0 | 2.5 | 3.0 | 3.5 | 4.0 | 4.5 | 5.0 |
| 500-Kb | Raw | 19 | 21 | 22 | 22 | 22 | 22 | 22 | 22 | 22 |
| | Normalized | 19 | 20 | 21 | 21 | 21 | 21 | 21 | 21 | 21 |
| 1.0-Mb | Raw | 18 | 19 | 20 | 21 | 21 | 21 | 21 | 21 | 21 |
| | Normalized | 18 | 19 | 20 | 20 | 20 | 20 | 20 | 20 | 20 |
| 1.5-Mb | Raw | 18 | 19 | 19 | 19 | 19 | 19 | 19 | 19 | 19 |
| | Normalized | 18 | 19 | 19 | 19 | 19 | 19 | 19 | 19 | 19 |
| 2.0-Mb | Raw | 17 | 18 | 18 | 18 | 18 | 18 | 18 | 18 | 18 |
| | Normalized | 17 | 18 | 18 | 18 | 18 | 18 | 18 | 18 | 18 |
| 2.5-Mb | Raw | 17 | 18 | 18 | 18 | 18 | 18 | 18 | 18 | 18 |
| | Normalized | 17 | 18 | 18 | 18 | 18 | 18 | 18 | 18 | 18 |

Note: This is the summary of all the cluster results from a standard EndHiC run for the great burdock, with contig end sizes from 500-Kb to 2.5-Mb by the step of 500-Kb, and contact cutoffs from 1 to 5 times of the turning point by the step of 0.5 times, using both raw and normalized Hi-C matrix data from HiC-Pro.

**Table S2.** Final merged cluster results of EndHiC, with robustness information.

| Cluster_id | Contig count | Cluster length (bp) | Robustness [max:90] | Included contigs with order and orientation |
|---|---|---|---|---|
| Cluster_A01 | 1 | 180,134,246 | 90 | ptg000007l+ |
| Cluster_A02 | 4 | 157,781,296 | 54 | ptg000019l+;ptg000017l+;ptg000037l+;ptg000033l- |
| Cluster_A03 | 2 | 137,502,900 | 81 | ptg000003l+;ptg000004l- |
| Cluster_A04 | 2 | 116,070,885 | 80 | ptg000002l+;ptg000026l+ |
| Cluster_A05 | 1 | 103,325,058 | 90 | ptg000009l+ |
| Cluster_A06 | 1 | 98,139,150 | 90 | ptg000005l+ |
| Cluster_A07 | 2 | 91,022,879 | 90 | ptg000020l+;ptg000021l- |
| Cluster_A08 | 2 | 90,391,671 | 90 | ptg000011l-;ptg000023l+ |
| Cluster_A09 | 2 | 89,788,315 | 76 | ptg000018l-;ptg000028l+ |
| Cluster_A10 | 2 | 83,430,363 | 89 | ptg000010l-;ptg000025l+ |
| Cluster_A11 | 1 | 76,614,701 | 90 | ptg000013l+ |
| Cluster_A12 | 1 | 76,000,989 | 90 | ptg000012l+ |
| Cluster_A13 | 1 | 73,598,797 | 90 | ptg000022l+ |
| Cluster_A14 | 1 | 70,868,591 | 90 | ptg000016l+ |
| Cluster_A15 | 1 | 69,740,074 | 90 | ptg000001l+ |
| Cluster_A16 | 2 | 67,669,029 | 90 | ptg000015l+;ptg000027l- |
| Cluster_A17 | 1 | 61,985,296 | 90 | ptg000008l+ |
| Cluster_A18 | 1 | 60,844,256 | 90 | ptg000014l+ |

Note: this is the final merged cluster results from a standard EndHiC run for the great burdock, each cluster corresponding to a natural chromosome. Robustness represents the frequency of this cluster type in the results of all parameter combinations. For the order and orientation of contigs, "+" stands for the forward strand, while "-" stands for the reverse strand. "ptg000003l+;ptg000004l-" is equal to "ptg000004l+;ptg000003l-".

**Table S3.** Summary of simulated and real contig data

| Assembly type | Species | Chromo-somes (n) | Genome size | Total contig number | Total contig length | Assembled percent (%) | Average contig number | Contig N50 | Contig N90 |
|---|---|---|---|---|---|---|---|---|---|
| Simulated assembly | Human | 23 | 3,054,815,472 | 104 | 3,054,815,472 | 100% | 4.52 | 47,909,438 | 14,532,355 |
| | Rice | 12 | 373,094,580 | 42 | 373,094,580 | 100% | 3.50 | 11,071,427 | 5,000,429 |
| | Arabidopsis | 5 | 119,146,348 | 11 | 119,146,348 | 100% | 2.20 | 9,660,775 | 5,994,203 |
| hifiasm-assembly | Human | 23 | 3,054,815,472 | 82 | 3,007,080,905 | 98% | 3.57 | 89,131,734 | 28,203,557 |
| | Great burdock | 18 | 1,720,000,000 | 30 | 1,709,056,189 | 99% | 1.67 | 74,692,580 | 38,981,084 |
| | Water spinach | 15 | 485,000,000 | 29 | 480,197,403 | 99% | 1.61 | 23,511,778 | 9,860,712 |

Note: Average contig number, means average contig number per chromosome. Contigs with size > 1 Mb are used for statistics. For simulated data, the reference genomes of human CHM13 v1.1, rice (Nipponbare) ASM386523v1, Arabidopsis thaliana (Columbia) TAIR10.1 were used to simulate large contigs, and each chromosome of reference genome was randomly split into 1 to 6 contigs. For real data, the hifiasm-assembled contigs of human were downloaded from https://zenodo.org/record/4393631/files/CHM13.HiFi.hifiasm-0.12.fa.gz , while the contigs of great burdock and water spinach were assembled by hifiasm using default parameters from HiFi reads downloaded from NCBI-SRA databases (PRJNA764011 and PRJNA764042).

**Table S4.** Summary of Hi-C data and HiC-Pro results

| Species | Total bases of Hi-C data | Coverage depth (X) | Total read pairs | Uniquely aligned pairs | Self-circle | Dangling-end | Valid pairs | Valid pairs (duplicate removed) | Useful data percent |
|---|---|---|---|---|---|---|---|---|---|
| Human | 124,972,344,900 | 41 | 416,574,483 | 124,673,886 | 33,998 | 2,938,444 | 110,014,494 | 98,391,593 | 23.6% |
| Rice | 21,470,420,100 | 58 | 71,568,067 | 45,777,278 | 34,487 | 79,982 | 45,002,791 | 40,821,917 | 57.0% |
| Arabidopsis | 10,288,604,600 | 86 | 102,886,046 | 45,244,994 | 978,212 | 381,050 | 37,791,223 | 21,220,262 | 20.6% |
| Great burdock | 201,613,850,100 | 117 | 672,046,167 | 269,305,617 | 134,310 | 93,807 | 268,112,029 | 212,922,075 | 31.7% |
| Water spinach | 64,406,145,600 | 133 | 214,687,152 | 102,797,519 | 74,093 | 48,041 | 101,439,749 | 73,091,154 | 34.0% |

Note: The human Hi-C reads data was downloaded from https://github.com/marbl/CHM13, while the Hi-C reads data for rice, Arabidopsis, great burdock, and water spinach were downloaded from NCBI-SRA (SRR5748737, SRR681003, PRJNA764011 and PRJNA764042).

**Table S5.** Comparison of performance for 4 Hi-C scaffolding software

| Hi-C scaffolder | human | | | rice | | | Arabidopsis | | |
|---|---|---|---|---|---|---|---|---|---|
| | Elapsed time | CPU time | Peak Memory | Elapsed time | CPU time | Peak Memory | Elapsed time | CPU time | Peak Memory |
| EndHiC | 0.5 m | 3.6 m | 159.1 M | 0.2 m | 1.1 m | 127.0 M | 0.1 m | 0.6 m | 20.4 M |
| <span style="color:red">LACHESIS</span> | <span style="color:red">5.6 m</span> | <span style="color:red">5.6 m</span> | <span style="color:red">2.7 G</span> | <span style="color:red">9.9 m</span> | <span style="color:red">9.9 m</span> | <span style="color:red">378.4 M</span> | <span style="color:red">1.5 m</span> | <span style="color:red">1.5 m</span> | <span style="color:red">295.0 M</span> |
| ALLHiC | 22.3 m | 43.2 m | 38.2 G | 17.1 m | 40.7 m | 16.8 G | 3.2 m | 7.0 m | 2.8 G |
| 3D-DNA | 21.9 h | 35.0 h | 51.6 G | 5.9 h | 13.6 h | 51.6 G | 1.7 h | 8.9 h | 28.7 G |

Note: the simulated data of three model species were used to run EndHiC, LACHESIS, ALLHiC and 3D-DNA. LACHESIS (highlighted in red) only finished the cluster step and then exited, it did not perform order and orientation here. The reason might be that LACHESIS grouped all the simulated large contigs into a single cluster and produced invalid intermediate files for later steps. Based on previous experience, the speed and memory consumption of LACHESIS is comparable to that of ALLHiC.

**Table S6.** Comparison of clustering accuracy for 4 Hi-C scaffolding software

a. Statistics of assembled chromosomes using simulated data

| Hi-C scaffolder | human (n = 23) | | | rice (n = 12) | | | Arabidopsis (n = 5) | | |
| --- | --- | --- | --- | --- | --- | --- | --- | --- | --- |
| | Complete | Mis-joined | Segment | Complete | Mis-joined | Segment | Complete | Mis-joined | Segment |
| EndHiC | 23 | 0 | 0 | 12 | 0 | 0 | 5 | 0 | 0 |
| 3D-DNA | 12 | 0 | 11 | 5 | 7 | 0 | 4 | 0 | 1 |
| LACHESIS | 0 | 23 | 0 | 0 | 12 | 0 | 0 | 5 | 0 |
| ALLHiC | 0 | 23 | 0 | 0 | 12 | 0 | 0 | 5 | 0 |

b. Statistics of assembled chromosomes using real data

| Hi-C scaffolder | human (n = 23) | | | great burdock (n = 18) | | | water spinach (n = 15) | | |
| --- | --- | --- | --- | --- | --- | --- | --- | --- | --- |
| | Complete | Mis-joined | Segment | Complete | Mis-joined | Segment | Complete | Mis-joined | Segment |
| EndHiC | 23 | 0 | 0 | 18 | 0 | 0 | 15 | 0 | 0 |
| 3D-DNA | 13 | 0 | 10 | 14 | 4 | 0 | 3 | 12 | 0 |
| LACHESIS | 0 | 23 | 0 | 0 | 18 | 0 | 0 | 15 | 0 |
| ALLHiC | 14 | 7 | 2 | 0 | 12 | 6 | 0 | 15 | 0 |

Note: For human, rice, Arabidopisis, the reference genomes of human CHM13 v1.1, rice (Nipponbare) ASM386523v1, Arabidopsis thaliana (Columbia) TAIR10.1 were used for comparison with the results of four Hi-C scaffolders. For great burdock and water spinach, the genomes of closely related species *Cynara cardunculus* (NCBI RefSeq GCF_001531365.1) and *Ipomoea triloba* (NCBI RefSeq GCF_003576645.1) were used for comparison.

**n**: haploid chromosome number of an organism.

**Complete**: A Hi-C scaffold includes most fragments only from one chromosome, i.e. the total length of shared contigs between a Hi-C scaffold (cluster) and the reference chromosome accounts for over 85% of both the scaffold (cluster) and the chromosome.

**Mis-joined**: A Hi-C scaffold includes large fragments from two or more chromosomes.

**Segment**: A chromosome is fragmented into two or more large scaffolds.